# Nanoseconds field emitted current pulses from ZrC needles and field emitter arrays

R. Ganter, R.J. Bakker, R. Betemps, M. Dehler, T. Gerber, J. Gobrecht, C. Gough, M. Johnson, E. Kirk, G. Knopp, F. Le Pimpec, K. Li, M. Paraliev, M. Pedrozzi, L. Rivkin, H. Sehr, L. Schulz and A. Wrulich

*Paul Scherrer Institute, WSLA-004, Villigen CH 5232, Switzerland,*

romain.ganter@psi.ch

Abstract

The properties of the electron source define the ultimate limit of the beam quality in linear accelerators like Free Electron Lasers (FEL). The goal is to develop an electron gun delivering beam emittance lower than current state of the art. Such a gun should reduce the cost and size of an X-ray Free Electron Laser (XFEL). In this paper we present two concepts of field emitter cathodes which could potentially produce low emittance beam. The first challenging parameter for such cathode is to emit peak current as high as 5 A. This is the minimum current requirement for the XFEL concept from Paul Scherrer Institut.[1] Maximum current of 0.12 A and 0.58 A have been reached respectively with field emitter arrays (FEA) and single needle cathodes. Laser assisted field emission gave encouraging results to reach even higher peak current and to pre-bunch the beam.

**Introduction**

The design of an electron gun capable of producing beams with an emittance one order of magnitude lower than current technology [2,3] would reduce considerably the cost and size of a XFEL radiating at 0.1nm.[1] In addition to low beam emittance, the electron source should also deliver peak currents of several amperes in order to drive efficiently a XFEL. Worldwide electron gun development is concentrated mainly on photocathodes[4,5] or thermionic cathodes[6] for such accelerator applications. An alternative gun design could be based on field emission.[7] Two kinds of field emitter sources are currently explored: field emitter arrays (FEAs) and single tip (needle) cathodes. With FEAs it is foreseen to use two gate layers above the matrix of tips: the first gate extracts the electrons from each tip and the second gate focuses them. This should produce an electron beam having the diameter of the FEA (below one millimeter) with very small divergence (controlled by the focusing gate layer). In the case of a needle cathode, the initial beam size is extremely small (in the order of a few micrometers) and the beam divergence is defined by electric field lines (which depend on the geometrical shape around the emitting area). Table 1 summarizes the targeted parameters for the two types of cathodes.

The first challenge is to achieve several amperes of peak current without tip destruction. Great care in handling and processing the cathodes together with emission duration in the nanosecond region at low repetition rate (10Hz) gave encouraging results to reach high peak current emission.[8] In this paper, we present the maximum peak current reached with FEAs and needle cathode as well as new perspectives to pre-bunch the beam with a laser.

**Limitations to high peak current emission**

The basic limitation to high current emission from sharp tips is the generation of vacuum arcs between tips and surrounding materials. The interaction between the electron beam and the surrounding surfaces (gate layer, collector, vacuum chamber, …) induces desorption of gas which can be ionized and back bombard the tip surface. These ions can damage or contaminate the emitter and in worse cases, they can trigger a vacuum arc. In addition, the heat that is produced on tip surface due to Joule effect and/or Nottingham effect can also induce thermal desorption of neutral particles. Fortunately in a XFEL application, we are only interested in very short emission durations, ideally 15 ps at low repetition rate (< 100 Hz). Such operation regime should prevent the tip from overheating so that higher current densities could be reached [9]. Another limit to high current emission for FEAs is the tip to tip emission uniformity. A maximum number of tips must contribute to the emission in order to reach a high peak current (and also to obtain good beam uniformity).[10] Pulse conditioning techniques are currently under development to improve the uniformity.[11] Finally, current limitation due to space charge effect can occur when the accelerating field is not high enough (see below). The targeted accelerating field (see Table 1) should avoid this limitation.

**FEA electron sources**

In order to characterize FEA properties we tested commercially available samples from SRI Inc.[12] (Menlo Park, CA). Those FEAs have 50'000 molybdenum tips on a one millimeter diameter disk area. Tips are grown on a p-doped Si wafer with a resistivity around 10 – 30 Ω.cm and separated from a Mo gate layer by a 1 micrometer thick dielectric layer. The capacitance due to overlapping electrodes has been measured to be around 150 pF. This gives already a maximum cut-off frequency of the FEA around 20 MHz (assuming 50 Ω as for the characteristic impedance of the transmission line).

FEAs are tested in a triode configuration as represented in Figure 1. The FEA gate is grounded and negative voltage pulses with amplitude $V_{ge}$ and duration $\tau$ (full width at half maximum) are applied to the tips at a repetition rate of 10Hz. The typical rise time of voltage pulses is around 3 ns for 150V amplitude. A faraday cup is facing the FEA at a distance $d_{FEA-FC}$ variable from 3mm to 2cm. The faraday cup is continuously biased with a positive voltage $V_{FC}$. Figure 2 shows current pulses obtained by applying voltage pulses with $\tau$ =10 ns. These voltage pulses are not fully transmitted from the pulser to the tips because of the resistive and capacitive load of the FEA (the charging time of the FEA is longer than 10ns). In consequence the voltage amplitude effectively seen by the tips is only a fraction of the applied amplitude indicated on Figure 2. In order to avoid this filtering effect, we used pulses with $\tau$ =30 ns so that the effective voltages between tips and gate has enough time to reach the applied amplitude. Figure 3 represents the collected peak current $I_{coll}$ versus applied amplitude $V_{ge}$ when $\tau$ = 30 ns, that is to say the current – voltage characteristic of the FEA.

Together with pulse conditioning techniques,[13] peak current as high as 120 mA have been measured. With such emitted current level, the resistivity of the silicon wafer as well as space charge effects begin to play a significant role. Figure 3 illustrates the influence of the electric field which accelerates electrons between the FEA and the faraday cup. In Figure 3 two sets of measurements have been taken with $d_{FEA-FC}$ = 2cm and $d_{FEA-FC}$ = 3mm respectively with a fixed $V_{FC}$ equal to 4.5kV. This leads to two mean accelerating fields ($E_{acc.}$) in the gap between the FEA and the faraday cup ($E_{acc.}=V_{FC} / d_{FEA-FC}$ ). With only 0.2 MV/m of accelerating field the emission saturates because of space charge effects [14]. At 1.5 MV/m, the influence of space charge is less important, however a Fowler - Nordheim (FN) plot (see inset of Figure 2) shows again a departure from the expected straight line (around $1/V_{ge} \sim 0.01$).

Accelerating gradients in electron guns are in general much larger (several hundreds of MV / m) so that our targeted peak current should not be limited by space charge effect.

These tests enabled us to better identify what would be the ideal FEA for our gun application. We initiated a program of FEAs development and the so-called molding technique[15,16] has been chosen as a fabrication technology. It is a more simple procedure in comparison to some other FEA production techniques (e.g. Spindt method) but it leads to less sharp emitters and requires higher tip to gate voltage. Figure 4 represents pictures of a pyramidal tip array and the top view of a single gated tip. In our application, such a pyramidal tip shape with large apex radius will help to dissipate heat when large current is emitted. The wafer which supports the tips is metallic (no silicon) so that material resistance is smaller than in commercially available FEAs.

In order to modulate the emission to picosecond level we will adopt a low capacitance FEA design similar to Ref [17]. In parallel, we are currently exploring another way to produce short electron bunches by using picosecond laser pulses to assist field emission.

**Single tip electron sources**

Single needle cathodes represent an interesting candidate for low emittance electron sources. These single needle cathodes are made from etched wires and can be coated and formed in order to carry high current emission. Such single tip requires a high voltage (tens of kilovolts) to emit since there is no close spaced gate layer. Standard electronic devices cannot produce kilovolt pulses of 15 ps duration. One way to achieve short emission duration is to use pulsed laser light illuminating the tip while high electric field is applied as in [18] and [19]. The high electric field present around the tip apex lowers the potential barrier of the crystal so that photoemission with less energetic photons becomes possible. This process presents several advantages for a low emittance gun application: 1) very short emission duration with picosecond laser pulses, 2) no UV light is required to reach high quantum efficiency and 3) photoelectrons are emitted with less initial kinetic energy (so less transverse kinetic energy which increases emittance) than with standard photocathode.

Tests have been carried out with a Nd-YAG laser delivering pulses of duration $\tau_{laser}$ ($\tau_{laser}$ = 10ns full width at half maximum) at 532 nm and illuminating a ZrC tip from AP Tech Inc.[20]. The challenge is to extract a minimum of 5 A from the apex of a needle tip (typically an area of a few micrometer square) without destroying it. Figure 5 represents the experimental setup. The tip faces a faraday cup which is at 2 cm distance. Negative voltage pulses (200ns to 1 ms at 20 Hz) can be applied to the tip with maximum amplitude of – 5 kV. The Faraday cup can be biased with a DC voltage of up to 25 kV. The laser beam is perpendicular to the tip axis. Figure 6 represents the typical laser intensity profile and the corresponding current pulse collected on the faraday cup (laser pulse is delayed because of a longer transmission line). This current pulse has been measured with a bias T in series with the faraday cup. If we assume a laser beam of 5mm diameter and energy of 100 mJ per pulse, we can estimate that the peak power illuminating the tip ($P_{532nm}$) was around $10^8$ W/cm$^2$ (with $\tau_{laser}$ = 10ns). Without applying the voltage, no current is detected: photons at 532 nm (2.3 eV) cannot extract photoelectrons from ZrC which has a work function around 3.5 eV.[21] When the electric field is present around tip apex, the potential barrier is lowered and electrons excited by 532nm photons can escape into vacuum. The tip radius r is around one micrometer leading to an electric field at the tip of F ~ 2 GV/m at V = 10 kV and assuming for the field enhancement factor β= F/V that β ~ 1/(5.r).[22]

Figure 7 represents current pulses measured when laser light was focused on the tip apex, giving higher laser power density (the laser power density has not been directly measured but a rough estimation gives: $2.10^8$ W/cm$^2$ < $P_{532nm}$ < $10^{11}$ W/cm$^2$). Current pulses are measured with a fast Rogowski coil from Bergoz (0.5 V/A). Peak current as high as 580 mA has been obtained with focused laser light and 14 kV between the tip and the faraday cup. In order to avoid laser ablation of the tip, the laser power was decreased to a level close to the lasing threshold. This caused large amplitude fluctuations of the laser power, directly translated into current pulse fluctuations (examples at 10 and 14kV on figure 7). Although there are amplitude fluctuations, figure 7 shows a dependence of the current with the applied voltage.

This dependence appears more clearly on figure 8 where averaged peak current values (over 25 acquisitions) versus voltage are represented. Voltage pulses are applied to the tip with a fixed amplitude of -5 kV (two pulses durations are tested: 200ns and 1000ns) and only the DC voltage applied to the Faraday Cup is varied. When the laser is off only field emitted electrons are collected with an onset voltage around 10 kV. Such a turn on voltage for field emission is compatible with a tip apex radius around one micrometer. When laser light is present, the current is more than one thousand times larger. The photocurrent dependence on the applied voltage follows the relation between the potential barrier height and the electric field F. As described by Garcia [18], the number of electrons per unit volume which can be possibly excited above the potential barrier by incident photons of energy hv, can be expressed as follow:

$$\rho_e(F) = \frac{1}{3\pi^2}\left(\frac{2m}{\hbar^2}\right)^{3/2}\left[E_F^{3/2} - \left(E_F + \Phi - h\nu - \sqrt{\frac{q^3 F}{4\pi\varepsilon_0}}\right)^{3/2}\right] \quad (1)$$

where m is the electron mass, $\hbar$ the Planck constant, $E_F$ the Fermi level, $\Phi$ the work function, q the elementary charge and $\varepsilon_0$ the electric constant. Electron density of equation (1) can be converted into a current value: $I = \rho_e(F).S.L.q/\tau_{laser}$ where S is the emitting area on tip apex, L the penetration depth of laser light. Figure 9 represents a comparison between the measured curves of Figure 8 and the current deduced from equation (1). In order to plot the measured values in function of the surface electric field F, we assumed that $F=\beta.V$ with $\beta$ equal to 200000 $m^{-1}$ (tip radius of one micrometer as mentioned above) and V the voltage between tip and faraday cup. Figure 9 shows a good qualitative agreement between the photo-field emission theory and the measurements. Higher electric fields give larger potential barrier reduction and thus more energy states accessible for photoemission. The quantitative fit in figure 9 has been obtained by assuming an emitting area of $S=10\mu m^2$ which is compatible with the actual tip size.

**Conclusion**

FEAs with diameters below one millimeter and needle tips are two possible cathode types for a low emittance electron gun for an FEL application to generate 0.1 nm X-rays. The first requirement for such an application is that a minimum peak current of 5.5 A must be emitted. Nanoseconds long current pulses of 0.12 A and 0.58 A have been successfully extracted from FEAs and single tip cathodes respectively. This represents an encouraging step towards our goal of 5 A in 15ps. In order to bunch the electron emission down to the picoseconds region, laser assistance seems to be an attractive solution according to the results presented in this paper. Such a photo-field emitter source acts like a photocathode with very high quantum yield at long wavelength (>532nm).

|                                   | **FEA**      | **Needle**   |
|-----------------------------------|--------------|--------------|
| Minimum Peak Current (A)          | 5.5          | 5.5          |
| Normalized Slice Emittance (m.rad)| $5.10^{-8}$  | $5.10^{-8}$  |
| Pulse Duration, rms (ps)          | 15           | 15           |
| Beam Diameter, rms (μm)           | < 600        | < 5          |
| Energy Spread (eV)                | < 0.5        | < 0.5        |
| Initial Beam Divergence (degree)  | < 0.01       | <10          |
| Accelerating Gradient (MV/m)      | > 250        | > 250        |

Manuscript #30138 – Romain Ganter

Table 1

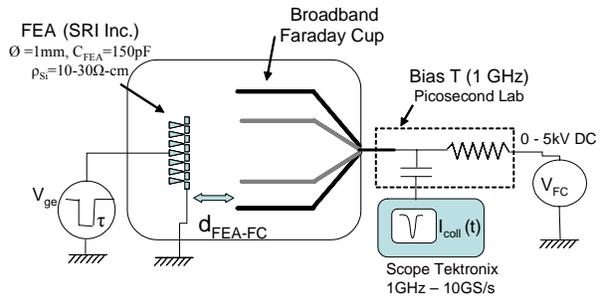

Figure 1

Manuscript #30138 – Romain Ganter

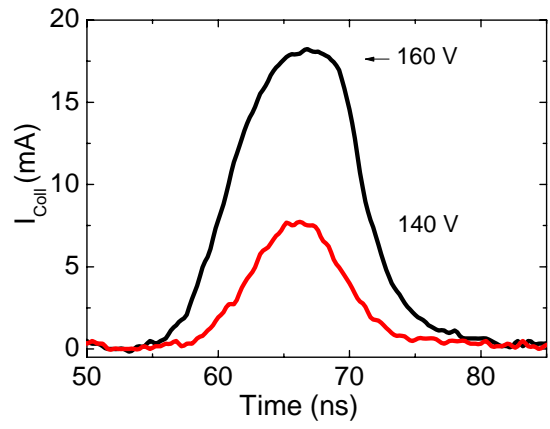

Figure 2

Manuscript #30138 – Romain Ganter

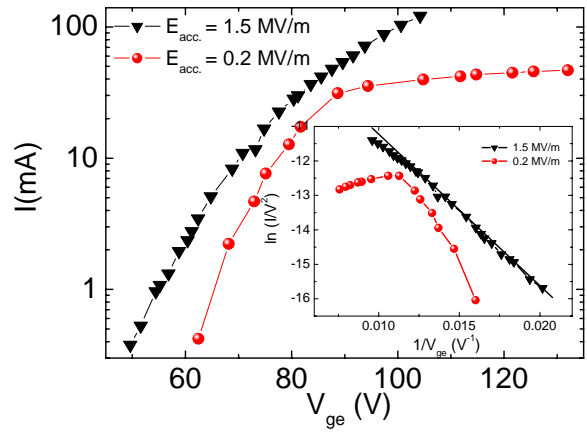

Figure 3

Manuscript #30138 – Romain Ganter

Figure 4

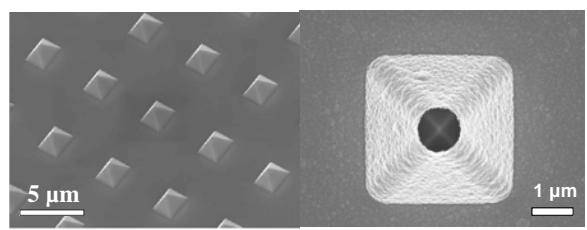

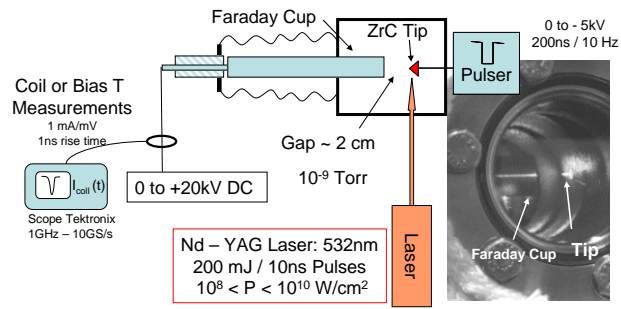

Figure 5

Manuscript #30138 – Romain Ganter

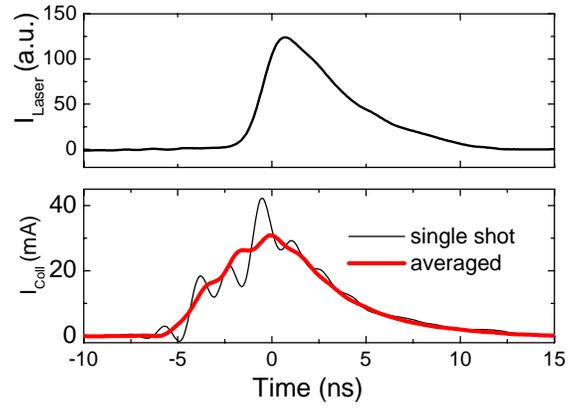

Figure 6

Manuscript #30138 – Romain Ganter

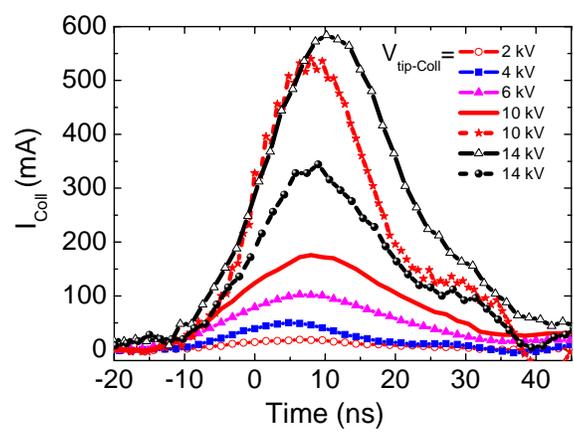

Figure 7

Manuscript #30138 – Romain Ganter

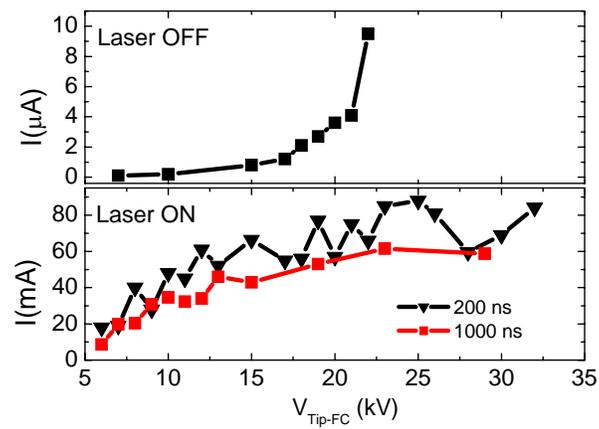

Figure 8

Manuscript #30138 – Romain Ganter

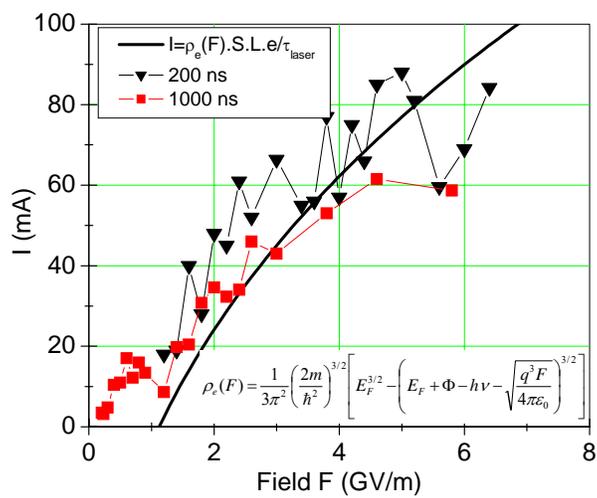

Figure 9

Manuscript #30138 – Romain Ganter

**List of Tables**

**List of Figures**

FIG. 1: Experimental setup used for measuring current voltage characteristics of commercial FEAs. $P \sim 10^{-9}$ torr

FIG. 2: Current pulses emitted by a FEA from SRI Inc. when $V_{ge}$=160V and $V_{ge}$=140V ($\tau$ = 10ns at 10Hz, $P \sim 10^{-9}$ torr, $d_{FEA-FC} \sim 2$ cm, $V_{FC}$ = 1.5 kV)

FIG 3: Current - voltage characteristics from a SRI FEA for two different mean accelerating field of 0.2 MV/m and 1.5 MV/m respectively. Inset: Fowler-Nordheim representation of the same curves. ($\tau$ = 30ns, $R_{FEA}$ is estimated $\sim 200\Omega$, $C_{FEA} \sim 150$pF, $P \sim 10^{-9}$ torr).

FIG 4: Array of molybdenum pyramidal tips (left) and top view of one gated pyramid (right).

FIG. 5: Experimental setup used for photo-field emission tests on a ZrC tip. Inset: Picture of the tip illuminated by unfocused laser light and facing the faraday cup.

FIG. 6: Laser intensity profile (top graph) and photo-field emitted current pulses (bottom graph). $V_{Tip}$= -5kV (Pulses of 1µs at 20 Hz), $V_{FC}$=1.5 kV, $P_{532nm} \sim 10^8$ W/cm$^2$, $10^{-9}$ torr.

FIG. 7: Photo-field emitted current pulses when laser light was focalized on tip apex ($\tau_{laser}$=10ns, $2.10^8 < P_{532nm} < 10^{11}$ W/cm$^2$, $V_{Tip-FC}$ represents the voltage between tip and faraday cup, gap $\sim$ 2cm, tip radius r $\sim$ 1µm, $10^{-9}$ torr).

FIG 8: Averaged peak current values versus Tip to Faraday Cup voltage V when laser is OFF (top graph) and when Laser is ON (bottom graph). $V_{Tip}$= -5kV (Pulses of 200ns and 1000ns at 20 Hz), $10^{-9}$ torr .

FIG 9: Comparison between measured photo-assisted current values (see Fig 8) and theoretical expression of the current. Tip radius: r=1μm, β=200000m$^{-1}$, $E_F$=7eV, Φ=3.5eV, hυ=2.3eV, emitting area S= 10 μm$^2$, laser penetration depth L=20nm, pulse duration $\tau_{laser}$=10ns, $10^{-9}$ torr .